\documentclass[12pt,english,a4paper]{article}
\usepackage[T1]{fontenc}
\usepackage[latin9]{inputenc}
\usepackage{graphicx}

\makeatletter

\usepackage{babel}
\makeatother

\begin{document}
\begin{center}
{\LARGE Exact entanglement entropy of the XYZ model and its sine-Gordon
limit}\\

\par\end{center}{\LARGE \par}

\begin{center}
{\large Elisa Ercolessi, Stefano Evangelisti
and Francesco Ravanini}\\

\par\end{center}{\large \par}

\begin{center}
Department of Physics, University of Bologna\\
and I.N.F.N., Sezione di Bologna\\
Via Irnerio 46, 40126 Bologna, Italy \\
ercolessi@bo.infn.it, stafano.evangelisti@gmail.com, ravanini@bo.infn.it\\
\par\end{center}

\begin{abstract}
We obtain the exact expression for the Von Neumann entropy for an
infinite bipartition of the XYZ model, by connecting its reduced density
matrix to the corner transfer matrix of the eight vertex model. Then
we consider the anisotropic scaling limit of the XYZ chain that yields
the 1+1 dimensional sine-Gordon model. We present the formula for
the entanglement entropy of the latter, which has the structure of
a dominant logarithmic term plus a constant, in agreement with what
is generally expected for a massive quantum field theory.\\

\end{abstract}

\noindent \textit{Classification codes} (PACS)\textit{\/}: {02.30.Ik, 11.10.-z,
75.10.Pq, 03.67.Mn, 11.10.-z}\\


\newpage{}

\section{Introduction}

Recently, the investigation of entanglement properties in systems
with many degrees of freedom has become a challenging line of research
which has put into evidence a growing number of connections between
quantum information and computational science, statistical mechanics,
quantum field theory as well as formerly very far away topics, such
as spin systems and black hole physics.

In particular, much attention has been devoted to the study of strongly
correlated spin and/or electronic models in one dimension, with the
aim to unveil the relationship between entanglement properties of
the ground state and quantum phase transitions. As soon as one has
to deal with systems made up of more than two two-level subsystems,
there does not exist a unique way of characterizing the degree of
entanglement stored by the system itself. Thus, several different
measures of entanglement have been proposed in literature over the
last few years and used to study the critical properties of the above
mentioned systems. These studies include more standard definitions
of entanglement measures, such as concurrence \cite{Fazio,Fazio2,Vidal},
as well as others, such as Renyi entropies \cite{Franchini}, local
\cite{Campos} indicators, or fidelity \cite{Zanardi}.

In the realm of quantum field theory, one usually studies entanglement
properties of a system via the computation of the so called Von Neumann
entropy $S=-\mbox{Tr}[\rho_{A}\log\rho_{A}]$ related to the reduced
density matrix $\rho_{A}$ of a subsystem A, as proposed by \cite{Bombelli},
 \cite{Callan} and \cite{Holzey} in the context of black hole physics, and
by \cite{Schumacher} in that of quantum information.
The general theory was then developed by Calabrese and 
Cardy \cite{Cardy}, in which  both the critical
(conformal) and the free massive cases have been examined. As for
the massless situation, much evidence has been accumulated proving
that, for a wide class of lattice systems, from the formula for $S$
one can extract the central charge of its scaling limit conformal
field theory. This is the case, for example, of spin $\kappa/2$ XXZ-chains
$(\kappa=1,2,3...)$ \cite{Weston} or of the $SU(3)$ AFM Heisenberg
chain \cite{Aguado}. In addition, some attention has been devoted
to the relationship between entanglement properties of the vacuum
and the boundary conditions of the theory \cite{Cardy,Weston,Asorey,Olalla}. Important
analytical progress has also been made in the context of massive integrable
quantum field models, for which a framework for the computation of
Von Neumann entropy has been developed using factorized scattering
and form factor techniques \cite{Doyon,Doyon2}.

In this paper we will first study the XYZ model and compute the exact
expression for the entanglement entropy for an infinite bipartition
of such quantum spin chain. As a measure of the entanglement, we will
use the Von Neumann entropy of the density matrix associated with
a semi-infinite segment of the chain. We will avoid the difficulties
concerning the direct computation of the density matrix by mapping
the quantum chain onto a two-dimensional classical spin system. As
Nishino \emph{et al.} \cite{Nishino1,Nishino2} first pointed out,
there is a connection between the density matrix of a quantum chain
and the classical partition function of a two-dimensional strip with
a cut perpendicular to it \cite{Peschel1}. In fact, the ground state
of the quantum chain described by a Hamiltonian $\hat{H}$ is also
an eigenvalue of the row-to-row transfer matrix $T$ of the classical
model, provided that $[\hat{H},T]=0$. In \cite{Cardy}, this analogy
was used in order to compute the entanglement of a transverse Ising
chain and of a XXZ chain near their critical points. By exploiting
the existing link between the eight vertex model and the XYZ chain
\cite{Sutherland}, we will obtain a formula for the entanglement
entropy for the latter, which once more confirms the following universal
expression for $S$, which is valid near criticality when the correlation
length $\xi$ is much larger than the lattice spacing $a$: \begin{equation}
S\sim(c/6)\log(\xi/a)+U\label{eq:intro1}\end{equation}
 where $c$ is the central charge of the conformal field theory describing
the critical point that the theory is approaching. 
This formula first
appeared in \cite{Cardy}, but it is a natural consequence of a similar conformal 
field theory formula first derived by \cite{Callan,Holzey}. A similar formula can 
also be found in \cite{Srednicki}, but without any mention of the central charge.

Here $U$ is a non-universal constant which is well known to depend
on the particular model under investigation. When the bipartition
is made up of two semi-infinite chains, as it is in this paper, it
has been noted by several authors \cite{Cardy,Weston,Asorey,Olalla}
that it contains information about the so-called Affleck-Ludwig boundary
entropy \cite{Affleck}.

At the present time, with the method of this paper, we are not able
to extract the boundary entropy information from such non-universal
$U$, a result that can be achieved only after having calculated the
Von Neumann entropy in the finite-length interval case, within the
same regularization scheme. Also, the study of the exact link between this
 term and the boundary
entropy needs more accurate calculations on correlation functions
that are not yet easily accessible in integrable models. The present
exact result, however, could become very useful the day one will be
able to compare it with calculations coming from independent methods.

Finally, we will be interested in a particular and relevant scaling
limit of the XYZ chain, which yields the 1+1 dimensional sine-Gordon
model \cite{Luther,McCoy}. After a brief discussion of the connection
between these two models in the thermodynamic limit, we will present
the formula for the entanglement entropy of the latter which has a
dominant logarithmic term in perfect agreement with what we expected
by seeing the sine-Gordon model as a perturbed $c=1$ conformal field
theory. This formula also gives an analytic expression for the constant
$U$.

\section{Entanglement entropy for the XYZ model via corner transfer matrix\label{sec:XYZ}}

Let us consider the quantum spin-$\frac{1}{2}$ XYZ chain, which is
described by the following hamiltonian\begin{equation}
\hat{H}_{XYZ}=-{\displaystyle \sum_{n}}(J_{x}\sigma_{n}^{x}\sigma_{n+1}^{x}+J_{y}\sigma_{n}^{y}\sigma_{n+1}^{y}+J_{z}\sigma_{n}^{z}\sigma_{n+1}^{z})\label{eq:XYZ1}\end{equation}
 where the $\sigma_{n}^{i}$ ($i=x,y,z$) are Pauli matrices acting
on the site $n$, the sum ranges over all sites $n$ of the chain
and the constants $J_{x}$, $J_{y}$ and $J_{z}$ take into account
the degree of anisotropy of the model. Without any loss of generality,
we can put $J_{x}=1$. In the following we will exploit the very well
known connection between the XYZ model and the eight vertex model
\cite{Baxter}. Indeed, as shown by Sutherland \cite{Sutherland},
when the coupling constants of the XYZ model are related to the parameters
$\Gamma$ and $\Delta$ of the eight vertex model %
\footnote{On the relation among $\Gamma,\Delta$ and the Boltzmann weights of
XYZ we adopt here the conventions of \cite{Baxter}. %
} at zero applied field by the relations\begin{equation}
J_{x}:J_{y}:J_{z}=1:\Gamma:\Delta\label{eq:XYZ4bis}\end{equation}
 the row-to-row transfer matrix $T$ of the latter model commutes
with the Hamiltonian $\hat{H}$ of the former. It is customary \cite{Baxter}
to parametrize the constants $\Delta$ and $\Gamma$ in terms of elliptic
functions\begin{equation}
\Gamma=\frac{1+k\mbox{sn}^{2}(i\lambda)}{1-k\;\mbox{sn}^{2}(i\lambda)}\qquad,\qquad\Delta=-\frac{\mbox{cn}(i\lambda)\mbox{dn}(i\lambda)}{1-k\;\mbox{sn}^{2}(i\lambda)}\label{eq:XYZ3bis}\end{equation}
 where $\mbox{sn}(x)$, $\mbox{cn}(x)$ and $\mbox{dn}(x)$ are Jacobian
elliptic functions while $\lambda$ and $k$ are parameters whose
domains are the following\begin{equation}
0<k<1\qquad,\qquad0<\lambda<I(k')\label{eq:XYZ4}\end{equation}
 $I(k')$ being the complete elliptic integral of the first kind of
argument $k'=\sqrt{1-k^{2}}$. We recall that this parametrization
is particularly suitable to describe the anti-ferroelectric phase
of the eight vertex model, that is when $\Delta<-1$, even if it can
be used in all cases by redefining the relations that hold between
$\Gamma$ and $\Delta$ and the Boltzmann weights.

Then, if relation (\ref{eq:XYZ4bis}) holds, according to \cite{Nishino1,Nishino2,Peschel1}, in the thermodynamic
limit may be obtained the reduced
density matrix $\hat{\rho}_{1}$ of the XYZ model relative to a semi-infinite
chain as product of the four Corner Transfer Matrices $\hat{A}=\hat{C}$
and $\hat{B}=\hat{D}$ of the eight vertex model at zero field. The
starting point is the density matrix defined by $\hat{\rho}=\mid0\,\rangle\langle\,0\mid$,
where $\mid0\,\rangle\in\mathcal{H}$ is the ground state in the infinite
tensor product space $\mathcal{H}=\mathcal{H}_{1}\otimes\mathcal{H}_{2}$
of our quantum spin chain. The subscripts $_{1}$ and $_{2}$ refer
to the semi-infinite left and right chains. $\hat{\rho}_{1}$ is then
defined as the partial trace: \begin{equation}
\hat{\rho}_{1}={\rm Tr}_{\mathcal{H}_{2}}(\hat{\rho})\label{added}\end{equation}
 More precisely one can write \begin{equation}
\hat{\rho}_{1}(\bar{\sigma},\bar{\sigma}')=(\hat{A}\hat{B}\hat{C}\hat{D})_{\bar{\sigma},\bar{\sigma}'}=(\hat{A}\hat{B})_{\bar{\sigma},\bar{\sigma}'}^{2}\label{eq:XYZ4a}\end{equation}
 where $\bar{\sigma}$ and $\bar{\sigma}'$ are particular spin configurations
of the semi-infinte chain. \\
 Generally speaking, the above quantity does not satisfy the constraint
${\rm Tr}\hat{\rho}_{1}=1$, so we must consider a normalized version
$\hat{\rho}_{1}'$. Let us definine for future convenience the partition
function $\mathcal{Z}={\rm Tr}\hat{\rho}_{1}$. For the zero field
eight vertex model with fixed boundary conditions, as it is shown
in \cite{Baxter,Baxter1,Baxter2}, one can write down an explicit
form of the Corner Transfer Matrices in the thermodynamic limit \begin{eqnarray}
\hat{A}_{d}(u) & = & \hat{C}_{d}(u)=\left(\begin{array}{cc}
1 & 0\\
0 & s\end{array}\right)\otimes\left(\begin{array}{cc}
1 & 0\\
0 & s^{2}\end{array}\right)\otimes\left(\begin{array}{cc}
1 & 0\\
0 & s^{3}\end{array}\right)\otimes...\nonumber \\
\hat{B}_{d}(u) & = & \hat{D}_{d}(u)=\left(\begin{array}{cc}
1 & 0\\
0 & t\end{array}\right)\otimes\left(\begin{array}{cc}
1 & 0\\
0 & t^{2}\end{array}\right)\otimes\left(\begin{array}{cc}
1 & 0\\
0 & t^{3}\end{array}\right)\otimes...\label{eq:XYZ5}\end{eqnarray}
 where $s$ and $t$ are functions of the parameters $\lambda$, $k$
and $u$ whose explicit expressions are given by $s=\exp\left[-\pi u/2I(k)\right]$
and $t=\exp\left[-\pi(\lambda-u)/2I(k)\right]$, $I(k)$ being an
elliptic integral of the first kind of modulus $k$. The downscripts $_{d}$
refer to a new basis in which the operators $A$ and $B$ are simultaneously
diagonal. We can indeed do that for the present purpose because we are 
interested in evaluating a trace, hence  we are only interested in the 
eigenvalues of $\rho_{1}$.
Thus in this new basis the density
operator of formula (\ref{eq:XYZ4a}) is given by\begin{equation}
\hat{\rho}_{1}=\left(\begin{array}{cc}
1 & 0\\
0 & x\end{array}\right)\otimes\left(\begin{array}{cc}
1 & 0\\
0 & x^{2}\end{array}\right)\otimes\left(\begin{array}{cc}
1 & 0\\
0 & x^{3}\end{array}\right)\otimes...\label{eq:XYZ6}\end{equation}
 where $x=(st)^{2}=\exp[-\pi\lambda/I(k)]$. We notice that $\hat{\rho}_{1}$
is a function of $\lambda$ and $k$ only and that, furthermore, it
can be rewritten as\begin{equation}
\hat{\rho}_{1}=(\hat{A}\hat{B})^{2}=e^{-\epsilon\hat{O}}\label{eq:XYZ7}\end{equation}
 $\hat{O}$ is an operator with integer eigenvalues and\begin{equation}
\epsilon=\pi\lambda/I(k)\label{eq:eps}\end{equation}
 The Von Neumann entropy $S$ can then be easily calculated according
to\begin{equation}
S=-\mbox{Tr}\hat{\rho}_{1}'\ln\hat{\rho}_{1}'=-\epsilon{\displaystyle \frac{\partial\ln\mathcal{Z}}{\partial\epsilon}}+\ln\mathcal{Z}\label{eq:XYZ8}\end{equation}
 where, in our case, the partition function is given by\begin{equation}
\mathcal{Z}={\displaystyle \prod_{j=1}^{\infty}}(1+x^{j})={\displaystyle \prod_{j=1}^{\infty}}(1+e^{-\pi\lambda j/I(k)})\label{eq:XYZ9}\end{equation}
 Thus we obtain an exact analytic expression for the entanglement
entropy of the XYZ model\begin{equation}
S_{XYZ}=\epsilon{\displaystyle \sum_{j=1}^{\infty}}{\displaystyle \frac{j}{(1+e^{j\epsilon})}}+{\displaystyle \sum_{j=1}^{\infty}}\ln(1+e^{-j\epsilon})\label{eq:XYZ10}\end{equation}
 which is valid for generic values of $\lambda$ and $k$. This is
the main result of this section. When $\epsilon\ll1$, i.e. in the
scaling limit analogous to the one of \cite{Weston}, formula (\ref{eq:XYZ10})
can be approximated by its Euler-Maclaurin asymptotic expansion, yielding\begin{eqnarray}
S_{XYZ} & = & {\displaystyle \int_{0}^{\infty}}\;{\rm d}x\;\left({\displaystyle \frac{x\epsilon}{1+e^{x\epsilon}}}+\ln(1+e^{-x\epsilon})\right)-{\displaystyle \frac{\ln2}{2}}+O(\epsilon)\nonumber \\
\, & = & {\displaystyle \frac{\pi^{2}}{6}}{\displaystyle \frac{1}{\epsilon}}-{\displaystyle \frac{\ln2}{2}}+O(\epsilon)\label{eq:XYZ11}\end{eqnarray}
 This will be used in the next subsection where we will check our
analytic results with some special known cases, namely the XXZ and
the XY chain.

\section{Some checks against known results}

Let us first consider the case $k=0$ (i.e. $\Gamma=1$) and the limit
$\lambda\to0^{+}$, which corresponds to the spin $1/2$ XXZ. In this
limit the eight vertex model reduces to the six vertex model. Let
us note that formula (\ref{eq:XYZ11}) coincides exactly with the
one proposed by Weston \cite{Weston} which was obtained in the study
of more general spin $\kappa/2$. In this limit the approximation
$\epsilon\ll1$ is still valid and we can use the result of equation
(\ref{eq:XYZ11}). From relation (\ref{eq:XYZ3bis}), it follows that
\begin{equation}
\lambda={\displaystyle \sqrt{2}}\sqrt{-\Delta-1}+O\left((-1-\Delta)^{3/2}\right)\label{eq:XYZ12}\end{equation}
 Thus equation (\ref{eq:XYZ11}) gives\begin{equation}
S=\frac{\pi^{2}}{12\sqrt{2}}\,\frac{1}{\sqrt{-\Delta-1}}-\frac{\ln(2)}{2}+O\left((-1-\Delta)^{1/2}\right)\label{eq:XYZ13}\end{equation}
 which can be written in a simpler form if we recall that, when $\epsilon\to0$
(i.e. $\Delta\rightarrow-1^{-}$), the correlation length is given
by \cite{Baxter} \begin{equation}
\ln\frac{\xi}{a}={\displaystyle {\frac{\pi^{2}}{\epsilon}}-2\ln(2)+O(e^{-\pi^{2}/\epsilon})}\end{equation}
 where $a$ is the lattice spacing. Recalling that \begin{equation}
\epsilon={\displaystyle 2\sqrt{2}}\sqrt{-\Delta-1}+O\left((-1-\Delta)^{3/2}\right)\end{equation}
 the expression for the entanglement entropy becomes \begin{equation}
S=\frac{1}{6}\ln\frac{\xi}{a}+U+O\left((-1-\Delta)^{1/2}\right)\end{equation}
 where $U=-\ln(2)/6$. This last expression confirms the general theory
of equation (\ref{eq:intro1}) with $c=1$, which is exactly what
one should expect, the XXZ model along its critical line being a free
massless bosonic field theory with $c=1$. The entanglement entropy in XXZ critical
chain was also derived from the second law of thermodynamics in \cite{Korepin3}.

\begin{figure}
\begin{centering}
\includegraphics[scale=0.7]{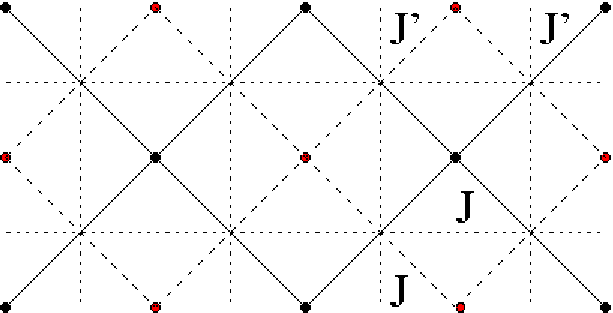} 
\par\end{centering}

\caption{Decoupled anisotropic Ising lattices. Horizontal and vertical lines
belong to the original eight vertex model lattice, diagonal lines
belong to the dual Ising lattice.}

\label{fig:ising} 
\end{figure}

As a second check, we consider the case $\Gamma=0$, which corresponds
to the XY chain. It is convenient now to describe the corresponding
eight vertex model by using Ising-like variables which are located
on the dual lattice \cite{Baxter}, thus obtaining an anisotropic
Ising lattice, rotated by $\pi/4$ with respect to the original one,
with interactions round the face with coupling constants $J,J'$,
as shown in figure \ref{fig:ising}. %

In our case the Ising lattice decouples into two single sublattices
with interactions among nearest neighbors. Now \begin{equation}
\Delta=\sinh(2\beta J)\sinh(2\beta J')\equiv k_{I}^{-1}\label{eq:XYZ17}\end{equation}
 so that, using the elliptic parametrization, one has\begin{equation}
\lambda=\frac{1}{2}I(k')\label{eq:XYZ18}\end{equation}
 Thus $\epsilon$ of equation (\ref{eq:eps}) becomes \begin{equation}
\epsilon=\frac{\pi I(k_{I}')}{I(k_{I})}\label{eq:XYZ19}\end{equation}
 Let us approach the critical line of the anisotropic Ising model
from the ferromagnetic phase, i.e. let us assume that $k_{I}\to1^{-}$.
In this case it is straightforward to write \begin{equation}
\epsilon=-{\displaystyle \frac{\pi^{2}}{\ln(1-k_{I})}}+O\left(\ln^{-2}(1-k_{I})\right)\label{eq:XYZ20}\end{equation}
 so that the entanglement entropy is\begin{equation}
S=-\frac{1}{6}\ln(1-k_{I})+O\left(\ln^{-1}(1-k_{I})\right)\label{eq:XYZ21}\end{equation}
 Since $\xi^{-1}=(1-k_{I})+O\left((1-k_{I})^{2}\right)$, we can easily
conclude that\begin{equation}
S=\frac{1}{6}\ln\frac{\xi}{a}+O\left(\ln^{-1}(1-k_{I})\right)\label{eq:XYZ22}\end{equation}
 where again the leading term confirms the general result (\ref{eq:intro1}),
with $c=1$. This result is in agreement with what found in previous
works \cite{Vidal,Cardy,Peschel3,Korepin, Korepin2} by means of different approaches.
In particular the results contained in \cite{Korepin, Korepin2} are rigorous mathematical theorems.

\section{The sine-Gordon limit}

In \cite{Luther,McCoy} it has been proposed that a particular scaling
limit of the XYZ model yields the sine-Gordon theory. In this section
we will use this connection to compute the exact entanglement entropy between
two semi-infinite intervals of a 1+1 dimensional sine-Gordon model.
In his article, Luther \cite{Luther} showed that in the scaling
limit, where $a\to0$ while keeping the mass gap constant, the parameters
of the XYZ model and those of the sine-Gordon theory are connected
by the following relation (keeping $J_{x}=1$ from the beginning)\begin{equation}
M=8\pi\,\left({\displaystyle \frac{\sin\mu}{\mu}}\right)\,\left({\displaystyle \frac{l_{r}}{4}}\right)^{\pi/\mu}\label{eq:SG1}\end{equation}
 where the parameter $\mu$ is defined as\begin{equation}
\mu\equiv\pi\left(1-{\displaystyle \frac{\beta^{2}}{8\pi}}\right)=\arccos\left({\displaystyle -J_{z}}\right)\label{eq:SG2}\end{equation}
 Here $M$ is the sine-Gordon solitonic mass, and $l_{r}=l\, a^{-\mu/\pi}$,
with\begin{equation}
l^{2}={\displaystyle \frac{1-J_{y}^{2}}{1-J_{z}^{2}}}\label{eq:SG3}\end{equation}
 These relations tell us how the coupling constant $J_{z}$ is connected
to the parameter $\beta$ of sine-Gordon, and how $J_{y}$ scales
when we take the scaling limit $a\to0$. It is clear from equation
(\ref{eq:SG3}) that in this limit $J_{y}\to1^{-}$. In the following
we work in the repulsive regime $4\pi<\beta^{2}<8\pi$ (which corresponds
to $0<\mu<\pi/2$ and $-1<J_{z}<0$). In this regime the mass gap
of the theory is the soliton mass $M$. Taking this limit we use the
following parametrization of the XYZ coupling constants\begin{equation}
\Gamma={\displaystyle \frac{J_{z}}{J_{y}}}\qquad,\qquad\Delta={\displaystyle \frac{1}{J_{y}}}\label{eq:SG4}\end{equation}
 which amounts to a reparametrization of the Boltzmann weights of
XYZ suitable for the $|\Delta|\leq1$ disordered regime where we are
working now (see chapter 10 of \cite{Baxter} for details). As a consequence
of such reparametrization a minus sign appears in front of both equations
(\ref{eq:XYZ3bis}). Taking the sine-Gordon limit, $\lambda$ and
$k$ parametrizing $\Gamma$ and $\Delta$ must now satisfy the following
constraint\begin{equation}
\mbox{sn}^{2}(i\lambda)=-{\displaystyle \frac{\frac{J_{z}}{J_{y}}+1}{k-k\frac{J_{z}}{J_{y}}}}\label{eq:SG5}\end{equation}
 Considering the parametrization of $\Delta$ and using the properties
of the Jacobian elliptic functions we can write\begin{equation}
\Delta^{2}=\frac{\mbox{cn}^{2}(i\lambda)\mbox{dn}^{2}(i\lambda)}{(1-k\;\mbox{sn}^{2}(i\lambda))^{2}}={\displaystyle \frac{\left(k(1-\frac{J_{z}}{J_{y}})+\frac{J_{z}}{J_{y}}+1\right ) \left(k(1+\frac{J_{z}}{J_{y}})-\frac{J_{z}}{J_{y}}+1\right)}{4k}}\label{eq:SG6}\end{equation}
 Expanding around $k\to1^{-}$ and collecting $\Delta^{2}=1/J_{y}^{2}$
from both sides of the equation we find\begin{equation}
\Delta^{2}=1+\frac{1}{4}(1-J_{z}^{2})(k-1)^{2}+O(k-1)^{3}\label{eq:SG7}\end{equation}
 Using equation (\ref{eq:SG1}) we obtain\begin{equation}
l^{2}=l_{r}^{2}a^{2\mu/\pi}=4^{2-3\mu/\pi}\left({\displaystyle \frac{M\mu a}{\pi\sin\mu}}\right)^{2\mu/\pi}\label{eq:SG8}\end{equation}
 where $\mu$ is completely fixed by choosing a particular value of
$J_{z}$. Now using the definition (\ref{eq:SG3}) and (\ref{eq:SG4})
we find\begin{equation}
\Delta^{2}=1+(1-J_{z}^{2})4^{2-3\mu/\pi}\left({\displaystyle \frac{M\mu a}{\pi\sin\mu}}\right)^{2\mu/\pi}+O(a^{4\mu/\pi})\label{eq:SG9}\end{equation}
 which is valid when $a\to0$. Comparing equation (\ref{eq:SG7})
with (\ref{eq:SG9}) we can identify in which way $k$ scales to $1^{-}$\begin{equation}
k=1-2^{3(1-\mu/\pi)}\left({\displaystyle \frac{M\mu a}{\pi\sin\mu}}\right)^{\mu/\pi}+O(a^{2\mu/\pi})\label{eq:SG10}\end{equation}
 Remembering the constraint (\ref{eq:SG5}) and using the previous
expression for $k$ we have\begin{equation}
\mbox{sn}^{2}(i\lambda)={\displaystyle \frac{-J_{z}-1}{1-J_{z}}+O(a^{\mu/\pi})}\label{eq:SG11}\end{equation}
 When $k\to1$ the elliptic function sn reduces to an hyperbolic tangent,
thus we obtain\begin{equation}
\tan^{2}\lambda={\displaystyle \frac{1+J_{z}}{1-J_{z}}+O(a^{\mu/\pi})\quad\longrightarrow\quad\lambda=\arctan\sqrt{{\displaystyle \frac{1+J_{z}}{1-J_{z}}}}+O(a^{\mu/\pi})}\label{eq:SG12}\end{equation}
 Now we can evaluate the expression (\ref{eq:XYZ10}) in this limit.
Using the following asymptotic behaviour of the elliptic integral
$I(x)$\begin{equation}
I(x)\approx-\frac{1}{2}\ln(1-x)+\frac{3}{2}\ln2+O(1-x),\qquad x\approx1^{-}\label{eq:SG13}\end{equation}
 along with the approximation (\ref{eq:XYZ11}), we can write the
exact entanglement entropy of a bipartite XYZ model in the sine-Gordon
limit\begin{equation}
S_{sG}=-{\displaystyle \frac{\pi}{12}\frac{\ln(1-k)-3\ln2}{\arctan\sqrt{{\displaystyle \frac{1+J_{z}}{1-J_{z}}}}}-{\displaystyle \frac{\ln2}{2}}+O(1/\ln(a))}\label{eq:SG14}\end{equation}
 The leading correction to this expression comes from the $O(\epsilon)$
term of equation (\ref{eq:XYZ11}). The constant $J_{z}$ is connected
to $\beta$ by\begin{equation}
J_{z}=-\cos\pi\left(1-\frac{\beta^{2}}{8\pi}\right)\label{eq:SG15}\end{equation}
 thus using this property and the scaling expression (\ref{eq:SG11})
we can write down the entanglement entropy as\begin{equation}
S_{sG}={\displaystyle \frac{1}{6}\ln\left(\frac{1}{Ma}\right)+\frac{1}{6}\ln\left(\frac{\sin\left[\pi\left(1-\frac{\beta^{2}}{8\pi}\right)\right]}{\left(1-\frac{\beta^{2}}{8\pi}\right)}\right)+O(1/\ln(a))}\label{eq:SG16}\end{equation}
 This result confirms the general theory due to \cite{Cardy,Doyon},
in the limit where the system is bipartite in two infinite intervals,
with the central charge equal to 1, as it should, because the sine-Gordon
model can be considered a perturbation of a $c=1$ conformal field
theory (described by a free massless boson compactified on a circle
of radius $\sqrt{\pi}/\beta$) by a relevant operator of (left) conformal
dimension $\beta^{2}/8\pi$. We can write \begin{equation}
S_{sG}\approx{\displaystyle \frac{1}{6}\ln\left(\frac{1}{Ma}\right)+U(\beta)\qquad a\to0}\label{eq:SG17}\end{equation}
 where the constant term $U(\beta)$ takes the value \begin{equation}
U(\beta)={\displaystyle \frac{1}{6}\ln\left(\frac{\sin\left[\pi\left(1-\frac{\beta^{2}}{8\pi}\right)\right]}{\left(1-\frac{\beta^{2}}{8\pi}\right)}\right)}\label{eq:SG18}\end{equation}

At $\beta^{2}=4\pi$, when the sine-Gordon model becomes the free
Dirac fermion theory, it assumes the value $U(\sqrt{4\pi})=\frac{1}{6}\log2=0.11552453...$,
while at $\beta^{2}=8\pi$, where the theory becomes a relevant perturbation
of the WZW conformal model of level 1 by its operator of left dimension
$\frac{1}{4}$, it becomes $U(\sqrt{8\pi})=\frac{1}{6}\log\pi=0.19078814...$.\\
 We notice that formula (\ref{eq:SG16}) yields the exact value of
the {\it overall} constant $U(\beta)$, since it has been derived from equation
(\ref{eq:XYZ11}) which is exact up to terms $O(\epsilon)$. As
mentioned in the introduction and as observed by many authors
\cite{Bombelli,Callan,Schumacher,Nishino2}, $U$ should contain a
contribution from the Affleck Ludwig boundary term as well as a
model-dependent constant that we are not able to extract at this
stage. Indeed, this would imply to obtain the Von Neumann entropy for
the finite-length interval within the same approach and by using the
same (lattice spacing) regularization scheme \footnote{This type of 
calculation can also shed some light on
understanding the universal part of the constant $U$
in the case of a massive feld theory, a result that should be
then compared with what is known in literature about the
conformal case.}.
In this respect it would be extremely interesting to be able to perform the same calculation
in other regularization schemes of the sine-Gordon model. Also, consideration
of boundary conditions other than the \char`\"{}vacuum\char`\"{} one
considered here could shed light, in the spirit of \cite{Weston},
on the problem of clearly relating this constant to the boundary entropy.
In this respect the considerations done in paper \cite{Olalla} are
very important. Clearly this question deserves further investigation
that we plan for the future.




\section{Conclusions and outlooks}

We have carried out an exact formula for the entanglement entropy
in an infinite bipartition of the XYZ model in the thermodynamic limit,
by which, as test bench, we have re-obtained some well known results
about the entanglement entropy of two integrable systems, the XXZ
model and the XY chain.

In addition we have obtained the entanglement entropy of the (repulsive)
sine-Gordon scaling limit of the XYZ model, which, on one side, confirms
once more the general theory due to \cite{Cardy,Doyon} and, on the
other side, yields for the first time an exact expression for such
quantity in the case of an interacting massive field theory. Since
we are dealing with a non free theory, an investigation of the connections
between this expression, Affleck-Ludwig boundary entropy \cite{Affleck}
and one point functions of sine-Gordon fields \cite{Zam} is non-trivial
and deserves further work.

Also it would be important to compare our results with those obtained
for massive 1+1 dimensional theories by the form factor approach of
\cite{Doyon,Doyon2} and generalized to any massive (even not integrable)
theory in \cite{Doyon3}. This implies to implement our technique
for a subsystem $A$ consisting of a finite interval, a situation
that we plan to study in the future.

Finally, let us remark that another issue addressed in \cite{Cardy}
is the dependence of entanglement entropy on finite size effects.
To investigate this aspect one should implement finite size effects
into the XYZ model, possibly within the Bethe ansatz approach, and
then rescale to the continuum. A good question to ask is how far this
approach can be related to the definition of sine-Gordon model on
a cylinder by rescaling lattice models, both in the XYZ approach \cite{Davide,Davide2}
or in the so called light-cone approach by Destri and De Vega \cite{DdV87}.
How far the finite size effects on entanglement entropy may be encoded
in structures generalizing the non-linear integral equations governing
e.g. the finite size rescaling of the central charge $c$ (see e.g.
\cite{Ravanini} and references therein) and the Affleck-Ludwig boundary
entropy $g$ \cite{Dorey,Lishman}, is an intriguing issue, surely
deserving further investigation.

All these lines of developments can shed new light on the comprehension
of the entanglement problem in quantum field theory.

\section*{Acknowledgments}

We wish to thank Francesco Buccheri, Andrea Cappelli, Filippo Colomo,
Marcello Dalmonte, Cristian Degli Esposti Boschi, Benjamin Doyon,
Davide Fioravanti, Federica Grestini and Fabio Ortolani for useful
and very pleasant discussions. This work was supported in part by
two INFN COM4 grants (FI11 and NA41) and by the Italian Ministry of
Education, University and Research grant PRIN-2007JHLPEZ.


\end{document}